\documentclass[12pt]{article}
\usepackage{epsf}
\usepackage{amsmath}
\usepackage{amsfonts}
\usepackage[mathscr]{euscript}
\begin{document}
\footskip50pt
\parindent=12pt
\parskip=.4cm
\input amssym.def
\input amssym
\def\tC{\tilde{C}}
\def\bR{{\Bbb R}}
\def\a{\"{a}}
\def\o{\"{o}}
\def\u{\"{u}}
\def\ep{\epsilon}
\def\eps{\varepsilon}
\def\l{\langle}
\def\r{\rangle}
\def\g{g^{\mu \nu}}
\def\d{{\rm d}}
\def\D{{\cal D}}
\def\L{{\cal L}}
\def\S{{\cal S}}
\def\J{{\cal J}}
\def\T{{\cal T}}
\def\X{{\cal X}}
\def\Y{{\cal Y}}
\def\ome{{\cal A}}
\def\om{{\omega}}
\def\Ga{\Gamma}
\def\pr{\partial}
\def\O{{\cal O}}
\def\R{{\cal R}}
\def\nab{\nabla}
\def\I{{\cal I}}
\def\J{{\cal J}}
\def\E{{\cal E}}
\def\hU{\hat{U}}
\def\hV{\hat{V}}
\def\hh{{1\over 2}}
\def\half{{\textstyle\frac{1}{2}}}
\def\quar{{\textstyle\frac{1}{4}}}
\def\ts{\textstyle}
\def\A{{\cal A}}
\def\B{{\cal B}}
\def\C{{\cal C}}
\def\de{\delta}
\def\si{\sigma}
\def\ga{\gamma}
\def\la{\lambda}
\def\ka{\kappa}
\def\be{\beta}
\def\al{\alpha}
\def\ga{\gamma}
\def\de{\delta}
\def\topcirc{\mathaccent"7017}
\def\oT{{\topcirc T}}
\def\oG{{\topcirc G}}
\def\oD{{\topcirc \D}}
\def\oDel{{\topcirc \Delta}}
\renewcommand{\theequation}{\thesection.\arabic{equation}}

\thispagestyle{empty}

\hfill hep-th/9905176

\vspace{1ex}

\begin{center}
{\Large \bf Gravitational Axial Anomaly for}

{\Large \bf Four  Dimensional Conformal Field Theories}  

{\Large \bf  }

\vspace{1.5cm}

{\parindent0cm
Johanna Erdmenger\footnote{Supported by Deutsche Forschungsgemeinschaft,
e-mail: Johanna.Erdmenger@itp.uni-leipzig.de; \newline address from 1st Sep
'99: Center for Theoretical Physics, MIT, Cambridge, MA 02139, USA}}

\vspace{1.5cm}

Institut f{\"u}r Theoretische Physik\\
Universit{\"a}t Leipzig\\
Augustusplatz 10/11\\
D - 04109 Leipzig\\
Germany
\end{center}

\vspace{2cm}

\centerline{\small \bf Abstract}
 
{ \small 
We construct the three point function involving an axial vector
current and two energy-momentum tensors for four dimensional conformal
field theories. Conformal symmetry determines the form of this three
point function uniquely up to a constant factor if the necessary conservation
conditions are imposed.   The gravi\-tational axial anomaly present
on a curved space background leads to a non-zero contribution for the
divergence of the axial current in this three point function even on 
flat space.
Using  techniques related to differential regularisation which guarantee
that the energy-momentum tensor is conserved and traceless,
we calculate the anomaly in the three point function directly. In this
way we relate the overall coefficient of the three point function 
to the scale of the gravitational axial anomaly. We apply our results
to the examples of the fermion and photon axial currents.
}

\vspace*{5mm}
\begin{tabbing}
\end{tabbing}
\newpage

\section{Introduction}

\setcounter{equation}{0}

Conformal symmetry constrains strongly
correlation functions in quantum field theory
not only in two, but also in higher dimensions. A
formalism for constructing conformally covariant  three point functions in general
dimensions $d$  involving
operators of arbitrary spin has been discussed in \cite{OP},
\cite{EO}. This formalism has been applied to three point functions which are
determined in terms of a finite number of linearly independent forms. In particular
expressions for conserved vector currents and the energy-momentum tensor were found
and various Ward identities were analysed.

Even for theories which are conformally invariant on flat space,
for a curved space background conformal invariance is broken by
anomalies involving the curvature for theories in even
dimensions. 
These anomalies generate a non-zero trace for the
energy-momentum tensor. On flat space they lead to
anomalous local contributions to the Ward identities reflecting
conformal symmetry. In consequence the coefficients of the conformal anomalies
are related to the coefficients of the independent forms of the
conformal correlation functions.
For example in two dimensions the central charge is present as 
coefficient of both 
the energy-momentum tensor two and three point functions and
the the conformal trace anomaly which in two dimensions is just the curvature
scalar. For the four dimensional case, the trace anomaly coefficients $c,a$
have been related to the coefficients of the three independent forms
present in general in the energy-momentum tensor three point function in \cite{EO}. 

Here we consider the conformal three point function $\l T_{\mu \nu}(x)
T_{\si \rho}(y) A_\om (z) \r$
involving an axial
vector current and two energy-momentum tensors in four dimensions. 
We show that conformal
invariance and energy-momentum conservation determine the structure of
this three point function uniquely up to an overall coefficient.
Moreover we relate this coefficient to the coefficient of the gravitational 
anomaly contributing to the divergence of the axial vector current on
a curved space background. This requires a careful analy\-sis of the
short-distance behaviour of the three point function, for which we
follow an approach analogous to the one we have used in \cite{EO} for
the axial anomaly contributing to the conformal three point function 
$\l V_\mu(x) V_\nu(y) A_\omega(z) \r$
involving two
conserved vector currents in addition to the axial current. The
relation between $\l V_\mu(x) V_\nu(y) A_\omega(z) \r$  and the corresponding
axial anomaly was first discussed in \cite{Schreier}.

The gravitational axial anomaly is of particular interest for
supersymmetric theories since supersymmetry relates the anomalous
divergence of the axial $R$ symmetry current to the anomalous trace of
the energy-momentum tensor depending on $c,a$. This anomaly has been used
to obtain formulae for the change in $c,a$ between fixed points \cite{Ccheck}
which strongly support the idea of a four dimensional extension of the 
Zamolodchikov 
C-theorem   \cite{Za}.

Recently, Pachos and Schiappa have also calculated contributions to the
conformal three point function $\l T_{\mu \nu}(x) T_{\si \rho}(y)
A_\om (z) \r$ in a second-order perturbative approach for the abelian
Higgs model \cite{PSc}. These authors claim to find two independent forms
in this correlation function. One of their expressions coincides with the
result found in this present paper, whereas the second form they discuss
does not appear to satisfy the necessary conservation condition for
the energy-momentum tensor.

This paper is organised as follows. We begin by constructing the
conformal three point function for an axial current and two 
energy-momentum tensors in four dimensions in section 2. In section 3 we
derive the anomalous Ward identity for this three point function for
quantum field theories coupled to a curved space background. By
analysing the short-distance behaviour of this three point function we
relate its coefficient to the coefficient of the gravitational axial
anomaly in section 4.  In section 5 we check our result by rederiving
the well-known result for the axial anomaly coefficient for free
fermions within our approach. As a second example we consider the 
case of the photon axial anomaly in section 6. Section 7 contains some concluding remarks.

\section{Conformal Three Point Function}

\setcounter{equation}{0}

We consider here four dimensional Euclidean space, although the formalism applies
in arbitrary dimensions. 
For three points $x,y,z$ we may  define a vector
$Z_\mu$ at $z$ by
\begin{gather} \label{Z}
Z_\mu = \half \pr^z {}_{\! \mu} \ln \frac{(z-y)^2}{(z-x)^2} = 
\frac{(x-z)_\mu }{(x-z)^2} - \frac{(y-z)_\mu }{(y-z)^2} \, , \quad
Z^2 = \frac{(x-y)^2 }{    (x-z)^2 (y-z)^2} \, ,
\end{gather}
which
transforms homogeneously under conformal transformations. 
$X_\mu$ and $Y_\mu$, which are vectors at $x$ and $y$, are obtained by
cyclic permutation, such that
\begin{gather} \label{X}
X_\mu = \frac{(y-x)_\mu}{(y-x)^2} - \frac{(z-x)_\mu}{(z-x)^2} \, ,
\quad
Y_\si = \frac{(z-y)_\si }{(z-y)^2} - \frac{(x-y)_\si}{(x-y)^2} \, .
\end{gather}
For constructing conformally invariant forms the inversion matrix $I$,
which is defined by
\begin{gather} \label{II}
I_{\mu \nu}(x) = \delta_{\mu \nu} - 2 \frac{x_\mu x_\nu}{x^2} \, ,
\qquad
{\rm det} I \, = \, -1 \, ,
\end{gather}
plays an essential role since for arbitrary conformal transformations,
$I$ acts as a parallel transport. 
The inversion matrix and the conformal vectors (\ref{Z}) satisfy the
identities
\begin{gather} 
I_{\mu\alpha}(x-z) Z_\alpha = - \frac{(x-y)^2}{ (z-y)^2} \, X_\mu \, , \quad
I_{\mu\alpha}(x-z)I_{\alpha\si}(z-y) = I_{\mu \si}(x-y) + 2 (x-y)^2
X_\mu Y_\si \, , \nonumber\\ 
I_{\si \al} (y-z) I_{\al \mu}(z-x) = I_{\si \nu} (y-x) I_{\nu \mu}(X)
\, , \label{Zid}
\end{gather}
which will prove to be very useful in the subsequent discussion.

For the construction of conformal three point functions in Euclidean space 
we use the formalism
developed in \cite{OP}, \cite{EO}, according to which conformal
covariance  restricts the three point function 
for quasi-primary operators $\O^i(x)$ of arbitrary spin to be of the form
\begin{gather}
\l \O_1^{i}(x) \, \O_2^{j} (y) \, \O_3^{k} (z) \r  \hspace{11cm} \nonumber\\
 = \frac{1}{(x-z)^{2\eta_1}\,(y-z)^{2\eta_2}} \,
D_1^{\, i} {}_{i'} (I(x-z))
D_2^{\, j} {}_{j'} (I(y-z)) \, t_{12,3}^{i'j'k} (Z) \, .  \label{3pt}
\end{gather}
Here $\eta_1$, $\eta_2$, $\eta_3$ denote the scale dimensions of the
operators $\O$. The indices $i,j,k$ denote components for the vector spaces
$V_1,V_2,V_3$ to which the fields $\O_1 , \O_2 , \O_3$ belong and 
which define representations of the group $O(4)$.
$D_1^{\, i} {}_{i'}(I)$ denotes the matrix acting on $V_1$ associated with an
inversion. In (\ref{3pt}) the tensor $t^{ijk}{}_{\!\! 12,3} (Z)$, belonging to
$V_1\otimes V_2 \otimes V_3$, is a homogeneous
function of the conformal vector $Z_\mu$ defined in (\ref{Z}),
\begin{align} \label{tijk}
t_{12,3}^{ijk}(\lambda Z) = {}& \lambda^{\eta_3 - \eta_1 - \eta_2}\,
t_{12,3}^{ijk}(Z) \, , \nonumber\\
D_1^{\, i} {}_{i'} (R) D_2^{\, j} {}_{j'} (R)
D_3^{\, k} {}_{ k'} (R) & \, t_{12,3}^{i'j'k'} (Z) =
t_{12,3}^{ijk}(RZ) \ \hbox {for all}\ R \in O(d) \, .
\end{align}
The general expression (\ref{3pt}) treats the three operators in an
equivalent way. To demonstrate this we note that (\ref{3pt}) may be
written equivalently in the form
\begin{gather} 
\l \O_1^{i}(x) \, \O_2^{j} (y) \, \O_3^{k} (z) \r  \hspace{10cm} \nonumber\\
= \frac{1}{ (x-y)^{2\eta_2}\,(x-z)^{2\eta_3}} \,
D_2^{\, j} {}_{j'} (I(y-x))
D_3^{\, k} {}_{k'} (I(z-x)) \, t_{23,1}^{j'k'i} (X) \, , \label{3ptX}
\end{gather}
where
\begin{gather}
t_{23,1}^{jk\, i} (X) = (X^2)^{\eta_1 - \eta_3} D_2^{\, j} {}_{j'} (I(X))\,
t_{12,3}^{ij'k}(- X)\, ,
\end{gather}
with $X$ as in (\ref{X}).

We now apply these general results to the three point function
involving the axial vector current $A_\omega$ as well as
two energy-momentum tensors $T_{\mu\nu}$. On flat space 
these operators satisfy
\begin{equation} \label{prop}
\pr_\omega A_\omega = 0 \, ,\quad \pr_\mu T_{\mu \nu} = 0 \, , \quad
T_{\mu \nu} = T_{\nu \mu} \, , \quad T_{\mu \mu} = 0 \, , 
\end{equation}  
discarding any
anomalies depending on the gauge fields
which do not contribute to the three point function
considered here.
In order to be conformal quasi-primary fields, $A_\om$ and $T_{\mu
  \nu}$ must have scale dimension $3$ and $4$ respectively. 
The general expression (\ref{3pt}) for the three point
function becomes in this special case 
\begin{gather} \label{tta}
\l T_{\mu \nu}(x) T_{\si \rho}(y) A_\omega(z) \r =
\frac{\I^T{}_{\!\!\mu \nu, \mu' \nu'} (x-z) \, \I^T{}_{\!\!\si \rho, \si' \rho'}
(y-z) }{(x-z)^8 (y-z)^8} \, t^{TTA}{}_{\mu' \nu' \si' \rho' \omega}(Z) \, ,
\end{gather}
where $\I^T$ is the inversion on $V_T$, the space of traceless
symmetric tensors,
\begin{gather}
\I^T{}_{\!\!\mu\nu,\si\rho}(x) = \E^T{}_{\!\!\!\! \mu \nu, \al \be} \, I_{\al
  \si}(x) I_{\be \rho}(x) \, , \\
\E^T{}_{\!\!\! \mu \nu, \al \be} = \half (\delta_{\mu \al} \delta_{\nu \be} +
\delta_{\mu \be} \delta_{\nu \al}) - \quar \delta_{\mu \nu} \delta_{\al
\be} \, 
\end{gather}
with $\E^T{}_{\!\!\!\mu \nu, \al \be}$
the projection operator onto $V_T$.
Bose symmetry imposes the condition
\begin{gather} \label{symm}
t^{TTA}{}_{\mu \nu \si \rho \omega}(Z) = 
t^{TTA}{}_{\si \rho \mu \nu \omega}(-Z) \, 
\end{gather}
on (\ref{tta}).
The most general solution of (\ref{symm}), which has also the correct 
dimension and reflects the symmetry and tracelessness properties of
the energy-momentum tensor and takes account of the odd parity
properties of the axial current, is given by
\begin{align} \label{ttta}
t^{TTA}{}_{\mu \nu \si \rho \omega}(Z) = \frac{1}{Z^6} \Big(
&  \A \, \E^T{}_{\!\!\! \mu \nu, \eta \eps} \,
\E^T{}_{\!\!\!\si \rho, \kappa \eps}
\, \eps_{\eta \kappa \omega \lambda} \, Z_\lambda  \, \nonumber\\
& {} + \B \,  \E^T{}_{\!\!\!\mu \nu, \eta \gamma} \, \E^T{}_{\!\!\!
\si \rho, \kappa \delta}
\, \eps_{\eta \kappa \omega \lambda} \, Z_\gamma 
Z_\delta Z_\lambda Z^{-2} \Big) \, ,
\end{align}
where $\A$, $\B$ are independent coefficients. 
This satisfies
\begin{equation}
\I^T{}_{\mu \nu, \mu' \nu'} (Z) \, \I^T{}_{\si \rho, \si' \rho'} (Z)
I_{\omega\omega'}(Z)\, t^{TTA}{}_{\mu' \nu' \si' \rho' \omega'}(Z)
=  t^{TTA}{}_{\mu' \nu' \si' \rho' \omega'}(-Z) \, ,
\end{equation}
as a consequence of
\begin{equation} \label{Iid}
I_{\mu \mu'} I_{\nu \nu' }I_{\al \al'} I_{\be \be'} \eps_{\mu' \nu'
  \al' \be'} = \det I \, \eps_{\mu \nu \al \be} = \, - \eps_{\mu \nu \al \be} \,  .
\end{equation}

Imposing
conservation of the energy-momentum tensor,
\begin{equation} \label{tcons}
\pr^x{}_{\!\!\mu}
 \l T_{\mu \nu}(x) T_{\si \rho}(y) A_\omega(z) \r \,= \,  0 \, , \qquad x\ne y,z \, ,
\end{equation}
implies the condition
\begin{gather} \label{ba}
\B \,  = \, - 6 \, \A
\end{gather}
on the parameters in (\ref{ttta}). Thus the three point function is
entirely determined by conformal invariance and by energy-momentum
conservation up to one overall
constant, such that there is only one independent form\footnote{In a
  recent paper by Pachos and Schiappa 
\cite{PSc} it was argued within a perturbative study of
 the abelian Higgs
  model that there are {\it two} independent forms in the three point
  function (\ref{tta}). However one of the two forms proposed by these
  authors coincides with the form obtained here, whereas the second
  does not satisfy the conservation condition (\ref{tcons}), (\ref{ba})
 for the energy-momentum tensor.}. Moreover the three point function
(\ref{tta}) automatically satisfies conservation with respect to the
axial vector leg for non-coincident points.

\section{Ward Identities}

\setcounter{equation}{0}

For massless quantum field theories on a curved space background 
in four dimensions the axial vector current may have an anomalous divergence
given by \cite{Eguchi} \cite{Salam} \cite{Witten}
\begin{equation} \label{cano}
\nabla_\omega \l A^\om (z) \r = \half c_A \, \eps_{\mu \nu \si \rho} R^{\mu
\nu\al \be} R^{\si \rho}{}_{ \al \be} \, .
\end{equation}
Here $R^{\mu}{}_{ \nu \si \rho}$ is the curvature tensor for the
background metric $g^{\mu \nu}$, and $c_A$ is a model-dependent coefficient. 

Furthermore on curved space the energy-momentum tensor may be defined by
\begin{gather} \label{ta}
\l T_{\mu \nu} (x) \r = \, - \, \frac{2}{\sqrt{g}} \, \frac{\delta
  W}{\delta g^{\mu \nu}(x)} \, , 
\end{gather}
with $W$ the generating functional for connected Green functions.
In consequence we have
\begin{gather} \label{3def}
\l T_{\mu \nu}(x) T_{\si \rho}(y) A_\omega (z) \r =
\frac{4}{\sqrt {g(x)} \sqrt {g(y)} \sqrt {g(z)}} \,  \frac{\delta}
{\delta g^{\mu \nu}(x)} \frac{\delta}{\delta g^{\si \rho}(y)}   
\Big ( \sqrt {g(z)} \l A_\om (z) \r \Big ) \, .
\end{gather}
Also the usual diffeomorphism invariance can here be expressed as
\begin{gather}\label{diff} 
\nabla^\mu \l T_{\mu \nu} (x) A_\om (z) \r = \nabla_\nu \Big( 
\de^{(4)}(x,z) g_{\mu\omega} \Big )  \l A^\mu (z) \r -
\nabla_\mu \Big( \de^{(4)}(x,z) g_{\nu\omega} \l A^\mu (z) \r \Big ) \, .
\end{gather}

Using (\ref{3def}) the anomalous divergence (\ref{cano})
leads to the anomalous Ward identity on flat space when 
$g_{\mu \nu} = \delta_{\mu \nu}$ and we may then identify up and down indices.
\begin{equation} \label{dtta}
\pr^z{}_{\!\omega} \l T_{\mu \nu} (x) T_{\si \rho} (y) A_\om (z) \r = 4
\, c_A \,
\E^T{}_{\!\!\!\mu \nu, \eta \eps} \, \E^T{}_{\!\!\! \si \rho, \kappa \eps} \,
\eps_{\eta \alpha \kappa \beta} \, \pr_\alpha \pr_\lambda \delta^{(4)}
(x-z) \pr_\beta \pr_\lambda \delta^{(4)}
(y-z) 
\end{equation}
for the three point function on flat space. Furthermore (\ref{diff}) has no anomaly
and (\ref{tcons}) must hold even for $x=y,z$.

In our discussion it is more convenient to replace the Riemann tensor by the
Weyl tensor defined by
\begin{equation}
C_{\mu \nu \si\rho} = R_{\mu \nu \si\rho} - (g_{\mu  [\si}
 R_{\rho] \nu} - g_{\nu [ \si} R_{\rho ] \mu } ) + 
{\ts
\frac{1}{3}} g_{\mu [\si} g_{\rho] \nu} R \, .
\end{equation}
The Weyl tensor belongs to the space $V^C = \{\C_{\mu \nu \si\rho}\}$ of tensors
defined by
\begin{equation} \label{Wsym}
\C_{\mu\nu\si\rho} =  \C_{[\mu\nu][\si\rho]} \, , \quad  \C_{\mu[\nu\si\rho]}
= 0 \, , \quad  g^{\mu\si}\C_{\mu\nu\si\rho} = 0 \, .
\end{equation}
It is important to recognise that
\begin{equation}
\C_{\mu\nu\si\rho} \in V^C \quad \Rightarrow \quad 
\C^*{}_{\! \mu\nu\si\rho} \equiv \half \eps_{\mu\nu\alpha\beta}\C^{\alpha\beta}
{}_{\si\rho} \in V^C \, .
\end{equation}
We may rewrite (\ref{cano}) as
\begin{equation} \label{cano2}
\nabla_\omega \l A^\om (z) \r =  c_A \, C^{\mu\nu\si\rho} 
C^*{}_{\! \mu\nu\si\rho} \, .
\end{equation}
If we define a projection operator $\E^C$ onto tensors in $V^C$ then 
to first order in an expansion about flat space, $ g_{\alpha\beta} = 
\delta _{\alpha\beta} + h_{\alpha\beta}$, the
the Weyl tensor is simply expressed as
\begin{gather} \label{ww}
C_{\mu\nu\si\rho} = 
2 \, \E^C{}_{\!\! \mu\nu\si\rho,\alpha\gamma\delta\beta}
\, \pr_\gamma \pr_\de \, h_{\alpha\beta} \, .
\end{gather}
An explicit expression for $\E^C$ may be found in \cite{EO}. 
With (\ref{ww})
 the flat space Ward identity (\ref{dtta}) may be rewritten equivalently as
\begin{equation} \label{dtta2}
\pr^z{}_{\! \omega} \l T_{\mu \nu} (x) T_{\al \be} (y) A_\om (z) \r =
16 \,  c_A \,
\E^C{}_{\!\!\mu \si \rho \nu, \al \ga \de' \be'}  \,  \,
\eps_{\de \be \de' \beta'} \,  \pr_\si \pr_\rho \delta^{(4)}
(x-z) \pr_\ga \pr_\de \delta^{(4)}
(y-z) \, .
\end{equation}

\section{Anomaly Calculation}

\setcounter{equation}{0}

Given the explicit form for the three point function provided by (\ref{tta}),
(\ref{ttta}) and (\ref{ba}) it is natural to endeavour to demonstrate
the explicit appearance of the anomaly as in (\ref{dtta2}) and obtain
$c_A$ explicitly in terms of $\A$. However a direct calculation does not give
a unique answer without further analysis
due to the very singular behaviour of the three point function for
$x\sim y \sim z$, due to which for instance its Fourier transform is
ill-defined. This singular behaviour
 may lead to anomalies in the conservation equations for the 
energy-momentum tensor in general. 
Therefore the axial anomaly is only unambiguous
when the energy-momentum tensor conservation equations are imposed 
even including contributions with support at $x=y=z$. To achieve this
we make use of the result that if, on flat space, 
\begin{equation} \label{tc}
T_{\mu\nu} = \pr_\si \pr_\rho \C_{\mu\si\rho\nu} \, , \qquad
\C_{\mu\si\rho\nu} \in V^C \, ,
\end{equation}
then $T_{\mu\nu}$ is automatically conserved and traceless.
Thus we write
\begin{equation} \label{ttcca}
\l T_{\mu \nu}(x) T_{\al \be}(y) A_\omega (z) \r =
\pr^x{}_{\!\si} \pr^x{}_{\! \rho} \pr^y{}_{\!\ga} \pr^y{}_{\! \de}
\Gamma^{CCA}{}_{\!\!\!\! \mu \si \rho\nu, \al \ga\de\be, \om}(x,y,z)
\, , 
\end{equation}
where $\Gamma^{CCA}(x,y,z)$ is a 
tensor belonging to $V^C \otimes V^C \otimes V$,
with $V$ the space of vector fields, $ A_\omega \in V$.
$\Gamma^{CCA}(x,y,z)$ is constructed according to the general prescription
for conformal three point functions regarding $\C_{\mu\si\rho\nu}$ as a
quasi-primary operator of dimension two. Writing the three point function in
the form (\ref{ttcca}) amounts to a form of differential regularisation 
\cite{diffreg}. A similar, albeit simpler, calculation of the 
axial anomaly
for a theory coupled to vector fields, maintaining the vector current
conservation equations was given in  \cite{EO}. For symmetry we must require
in (\ref{ttcca})
\begin{equation} \label{symm2}
\Gamma^{CCA}{}_{\!\!\!\! \mu \si \rho\nu, \al \ga\de\be, \om}(x,y,z)
= \Gamma^{CCA}{}_{\!\!\!\! \al \ga\de\be, \mu \si \rho \nu, \om}(y,x,z) \, ,
\end{equation}
and also for conservation of the axial current
\begin{equation}\label{con}
\pr^z{}_{\! \omega} \Gamma^{CCA}{}_{\!\!\!\! \mu \si \rho\nu, 
\al \ga\de\be, \om}(x,y,z) = 0 \, , \qquad z\ne x,y \, .
\end{equation}
A form for $\Gamma^{CCA}$ with the desired properties which has overall odd
parity is given by
\begin{gather} \label{cca1}
\Gamma^{CCA}{}_{\!\!\!\! \mu \si \rho\nu, \al \ga\de\be, \om}(x,y,z)
=  
{\ts \frac{1}{8}} \A \, \frac{\I^C_{\mu \si\rho \nu, \al \ga \de' \be'}(x-y)}{(x-y)^4 } 
\, \eps_{\de' \be' \de \be} \, Z^2 Z_\om \, , 
\end{gather}
where
\begin{gather}
\I^C{}_{\!\!\mu \si\rho \nu, \al \ga \de
    \be}(x) = \E^C{}_{\!\! \mu \si\rho \nu, \al' \ga' \de'
    \be'} I_{\al \al'}(x)   I_{\ga \ga'}(x)   I_{\de \de'}(x)   
I_{\be \be'}(x)  \, 
\end{gather}
is the representation of inversions on $V^C$. The symmetry constraint (\ref{symm2})
then follows from $\det I =-1$ once more.
The conformal three point function (\ref{cca1}) may also be written
in a form in which its consistency with the general formalism
(\ref{3pt}) is obvious,
\begin{gather}     
\Gamma^{CCA}{}_{\!\!\!\! \mu \si \rho\nu, \al \ga\de\be, \om}(x,y,z)
\hspace{11cm} \nonumber\\ =
\frac{ \I^C{}_{\!\! \mu \si \rho \nu, \mu' \si' \rho' \nu'}(
x-z) \I^C{}_{\!\! \al \ga\de\be,\al' \ga' \de' \be'}(
y-z)}{(x-z)^4 (y-z)^4} \, t^{CCA}{}_{\mu' \si' \rho' \nu', \al' \ga'
\de'  \be', \omega}(Z) \, , \label{cca2}
\end{gather}
with
\begin{gather}
t^{CCA}{}_{\mu \si \rho \nu,
\alpha\gamma\delta\beta, \omega}(Z) \hspace{11cm} \nonumber\\
= - { \ts \frac{1}{8}} \A \, \frac{Z_\om}{Z^2} \,  \E^C{}_{\!\! \mu \si \rho
 \nu, \mu' \si' \rho' \nu'} \E^C{}_{\!\! \al \ga \de \be, \al'  \ga' \de' \be'}
 \, \eps_{\de' \be' \kappa \lambda} \,  I_{\mu' \al'}(Z)
I_{\si' \ga'}(Z)I_{\rho' \kappa}(Z)I_{\nu' \lambda}(Z) \, , \label{ttc1}
\end{gather}
which may be obtained from (\ref{cca1}) by repeated use of
(\ref{Zid}) and (\ref{Iid}).  From the results of \cite{EO}
\begin{gather} \label{ttc2}
t^{TTA}{}_{\mu \nu \si \rho \omega}(Z) = \pr^Z{}_{\! \si} \pr^Z{}_{\! \rho}
\pr^Z{}_{\! \gamma} \pr^Z{}_{\! \delta}
t^{CCA}{}_{\mu \si \rho \nu, \alpha\gamma\delta\beta, \omega}(Z) \, .
\end{gather}
We have checked that inserting 
(\ref{ttc1}) into (\ref{ttc2}) 
and calculating the four derivatives 
using the algebraic computing program FORM \cite{FORM}, 
we obtain (\ref{tta}) with (\ref{ba}) as expected.

The expression in (\ref{cca1}) for $\Gamma^{CCA}(x,y,z)$ is homogeneous
of degree $-8$ in $x,y,z$. Any ambiguity is proportional to
$\delta^{(4)}(x-z) \delta^{(4)}(y-z)$, without derivatives, but the presence
of such a term is precluded by the tensorial structure.
Thus the divergence of the axial vector current in 
(\ref{con}) may be unambiguously calculated even including contributions
with support at $z=x,y$. For this purpose we
generalise results for distributions derived in \cite{EO}, which rely
on the general discussion of \cite{Gelfand}. 
First we use that
from (\ref{Z}) we have
\begin{gather}
\pr^z{}_\om (Z^2 Z_\om)  = 2 \pi^2 \left( 
\delta^{(4)}(y-z) - \delta^{(4)}(x-z)\right) \, .
\end{gather}
The remaining derivatives can be evaluated with the help of
\begin{gather}
\label{dist}
\pr^x{}_{\!\si} \pr^x{}_{\!\rho} \, 
\frac{\I^C{}_{\!\! \mu \si \rho \nu, \al \ga \de \be} (x) }{(x^2)^2}
=  { \frac{\pi^2}{6}} \, \E^C{}_{\!\! \mu \si \rho \nu, \al \ga \de \be}
  \pr^x{}_{\!\si} \pr^x{}_{\!\rho} \delta^{(4)}(x)
\end{gather}
in four dimensions. To obtain this we make use of the results for the singular
behaviour as $\lambda\to 0$
\begin{gather}
\pr^x{}_{\!\si} \pr^x{}_{\!\rho} \, \frac{\I^C{}_{\!\! \mu \si \rho \nu, \al \ga \de
\be} (x)
  }{(x^2)^{(4-\lambda)/2}} =  \frac{\lambda(\lambda+2)}{(4-\lambda)(6-\lambda)} \,
\E^C{}_{\!\! \mu\si\rho\nu,\al\ga\de\be} \, \pr^x{}_{\!\si} \pr^x{}_{\!\rho}
  \, \frac{1}{
(x^2)^{(4-\lambda)/2}} \nonumber\\ 
\hspace{1cm}{} \sim {\frac{\pi^2}{6}}\, 
\E^C{}_{\!\! \mu \si \rho \nu, \al \ga \de \be}
\,  \pr^x{}_{\!\si} \pr^x{}_{\!\rho} \, \delta^{(4)}(x) \,, 
\end{gather}
which depends on 
\begin{gather}
\frac{1}{(x^2)^{(4-\lambda)/2}} \sim  \frac{2 \pi^2}{\lambda} \, \de^{(4)}(x) \, .
\end{gather}
Furthermore in the same fashion for the singular term as $\lambda \to 0$
\begin{gather} 
\frac{\I^C{}_{\!\! \mu \si \rho \nu, \al \ga \de \be} (x)
  }{(x^2)^{(4-\lambda)/2}} \sim 0 \, ,
\end{gather}
so that the left hand side is well defined as a distribution at $\lambda =0$
and (\ref{dist}) is unambiguous.
Thus we obtain
\begin{equation} \label{dtta3}
\pr^z{}_{\!\omega} \l T_{\mu \nu} (x) T_{\al \be} (y) A_\om (z) \r = {\ts
 \frac{1}{12} } \pi^4 \A 
 \,  \,
\E^C{}_{\!\! \mu \si \rho \nu, \al \ga \de' \be'}  \,  \,
\eps_{\de \be \de' \beta'} \,  \pr_\si \pr_\rho \delta^{(4)}
(x-z) \, \pr_\ga \pr_\de \delta^{(4)}
(y-z) \, .
\end{equation}
Comparison with (\ref{dtta2}) yields
\begin{gather} \label{ca}
c_A \, = \, {\ts \frac{1}{192}} \pi^4 \A \, ,
\end{gather}
which relates the anomaly coefficient to the
coefficient of the conformal three point function (\ref{tta}). 

\section{Free Fermions}

\setcounter{equation}{0}

In order to check the result (\ref{ca}), we use it to rederive the
well-known result for $c_A$ in the case of free fermions, for which
\begin{gather}
T_{\mu \nu} = \bar \psi \ga_{(\mu} \! \overleftrightarrow {\pr}_{\!\!\nu)} \psi \,
, \quad  \overleftrightarrow \pr_{\!\!\nu} = \half ( \pr_\nu -
\overleftarrow {\pr}_{\!\!\nu}) \,,  \qquad A_\omega =  \bar \psi \ga_\omega \ga_5 \psi \, ,
\end{gather}
with conventions so that  ${\rm tr} (\gamma_\al \gamma_\be \gamma_\gamma \gamma_\de
\gamma_5) = 4 \, \eps_{\al \be \ga\de}$.
The basic two point function is
\begin{gather}
\l \psi(x) \bar \psi (0) \r = \frac{1}{2\pi^2} \frac{\gamma \cdot x}{x^4} 
\end{gather}
in four dimensions. This yields
\begin{gather}
\l T_{\mu \nu} (x) T_{\al \be} (y) A_\om (z) \r_\psi  \hspace{10cm}
\nonumber\\ = - \frac{1}{(2\pi^2)^3} \, {\rm tr} \, \Big[ \gamma_\omega \ga_5  
\frac{\gamma \cdot (z-x)}{(z-x)^4} \gamma_{(\mu} \! \overleftrightarrow
{\pr}^x{}_{\!\!\nu)}  \frac{\gamma \cdot (x-y)}{(x-y)^4} \ga_{(\si}
\! \overleftrightarrow {\pr}^y{}_{\!\!\rho)} \frac{\gamma \cdot (y-z)}{(y-z)^4} 
\nonumber\\ \hspace{4cm} {} + \gamma_\omega \ga_5  
\frac{\gamma \cdot (z-y)}{(z-y)^4} \gamma_{(\si} \! \overleftrightarrow
{\pr}^y{}_{\!\!\rho)}  \frac{\gamma \cdot (y-x)}{(y-x)^4} \ga_{(\mu}
\! \overleftrightarrow {\pr}^x{}_{\!\!\nu)} \frac{\gamma \cdot
  (x-z)}{(x-z)^4}\Big] \, . \label{fermions}
\end{gather}
This expression is most easily evaluated in a frame in which
$x,y,z$ are constrained to lie on a straight line. 
The result is of the form (\ref{tta}) with (\ref{ttta}) and satisfies the
conservation condition (\ref{ba}).
Comparison  of (\ref{fermions}) with (\ref{tta}) yields
\begin{gather}
\A =  \frac{1}{\pi^6}
\end{gather}
for free fermions,
which according to (\ref{ca}) corresponds to
\begin{gather}  
c_A = \frac{1}{192 \pi^2}
\end{gather}
for the fermion triangle anomaly . This agrees with the old result of Eguchi and
Freund \cite{Eguchi}, which provides a check on our
calculation. 

\section{Photons}
\setcounter{equation}{0}

A second field theory example in which there is an anomaly in the conservation of
an axial current for curved space backgrounds concerns the Chern-Simons current
in abelian gauge theories as described in \cite{Dolgov}.
The relevant current in terms of the gauge field is
\begin{gather} \label{Kom}
K_\omega = \eps_{\omega \al \be \ga} \, a_\al \, F_{\be \ga} \, , \qquad
F_{\be \ga} = \pr_\be a_\ga - \pr_\ga a_\be \, .
\end{gather}
Formally this satisfies
\begin{gather}
\pr_\omega K_\omega = F_{\be \ga}F^*{}_{\! \be \ga} \, ,
\label{KF}
\end{gather}
but there are potential anomalies arising from the photon triangle diagram.
The current $K_\omega$, defined in (\ref{Kom}) so as to satisfy (\ref{KF}), 
may be regarded as arbitrary up to 
\begin{gather} \label{Kg}
K_\omega \rightarrow K_\omega + \, \pr_\al W_{\omega \al} \, , \qquad
W_{\omega\alpha}  = - W_{\alpha\omega} \, .
\end{gather}
This allows for the freedom of gauge transformations since if
$a_\mu \rightarrow a_\mu + \pr_\mu \Phi$
the current transforms as in (\ref{Kg}) with
$W_{\omega\alpha} = \eps_{\omega \al \be \ga} \Phi F_{\be \ga}$.
 
For the gauge  field the gauge invariant energy-momentum tensor is given by
\begin{gather}
T_{\mu \nu} = F_{\mu \lambda} F_{\nu \lambda} - {\ts \frac{1}{4}}
\de_{\mu \nu} F_{\al \be} F_{\al \be} = \E^T{}_{\!\!\!\mu \nu, \mu' \nu'}
F_{\mu' \lambda} F_{\nu' \lambda} \, , \label{Tem}
\end{gather}
which is of course traceless and conserved subject to $\pr_\mu F_{\mu\nu}=0$.
In general with gauge fixing this need not be conserved even for free fields
but in a BRS formalism the full canonical conserved
energy momentum tensor, which need not be traceless,  may be written as
\begin{gather}
T^{\rm{can}}{}_{\!\!\! \mu\nu} = T_{\mu \nu} + s X_{\mu\nu} \, ,
\label{TX}
\end{gather}
where $s^2=0$. BRS invariance for a general three point function may be expressed
as
\begin{gather}
\label{BRS}
\l s A(x) \, B(y) \, C(z) \r + (-1)^{\varepsilon_A}\l  A(x) \, s B(y) \, C(z) \r 
+ (-1)^{\varepsilon_A + \varepsilon_B} \l A(x) \, B(y) \, s C(z) \r = 0 \, ,
\end{gather}
where $\varepsilon_A, \varepsilon_B$ are the BRS grading of the operators $A,B$.
For BRS invariant operators $s O_i = 0$ this shows that the three point function
is invariant under the freedom $O_i \to O_i + s X_i$.
In the following we endeavour to analyse the odd parity three point function
formed by two energy momentum tensors and the axial current $K_\om$
to see whether the
anomalous divergence suggested in \cite{Dolgov} can be derived in  a similar fashion
to the previous discussion for the axial fermion current. The current is not BRS
closed but in general we may write
\begin{gather}
s K_\omega =  \pr_\al W_{\omega \al} \, .
\end{gather}
In consequence it is easy to see using (\ref{BRS}) that we must have
\begin{gather}
\l T^{\rm{can}}{}_{\!\!\! \mu \nu} (x) T^{\rm{can}}{}_{\!\!\! \si \rho} (y) K_\om (z) \r
= \l T_{\mu \nu} (x) T_{\si \rho} (y) K_\om (z) \r 
+ \pr^z{}_{\! \al} W_{\mu\nu, \si \rho, \omega \al }(x,y,z)\, ,
\end{gather}
with $ W_{\mu\nu, \si \rho, \omega \al }(x,y,z) =
 W_{(\mu\nu), (\si\rho), \omega \al }(x,y,z) =  W_{\si\rho,\mu\nu,\omega \al }(y,x,z)$.
It is therefore sufficient to consider just
$\l T_{\mu \nu} (x) T_{\si \rho} (y) K_\om (z) \r $, although this need not
satisfy the energy momentum tensor conservation equations, so long as we allow for the
freedom represented by (\ref{Kg}) to obtain an expression satisfying these conditions.  
It is important to confirm whether this three
point function is BRS non trivial and hence cannot be transformed to zero.

In a standard covariant gauge the basic two point function for the gauge
field is given by
\begin{gather}
\l a_\mu (x) a_\si (y) \r_1 \, = \, { \frac{1}{8\pi^2}} \, \Big( (1+\al)
\frac{\de_{\mu \nu}}{(x-y)^2} + 2 (1-\al) \frac{(x-y)_\mu (x-y)_\nu}{(x-y)^4} \Big) \, ,
\label{aa}
\end{gather}
with $\al$ a gauge parameter, is not conformally covariant. However the two
point function for the gauge invariant field strength,
\begin{gather}
\l F_{\mu \nu}(x) F_{\si \rho}(y)  \r \, = \,  \frac{2}{\pi^2} \, \frac
{ \I^F{}_{\!\!\mu \nu, \si \rho}(x-y)}{(x-y)^4} \, , \label{FF}
\end{gather}
with
\begin{gather}
\I^F{}_{\!\!\mu\nu,\si \rho}(x-y) = \half ( 
I_{\mu \si}(x-y) I_{\nu \rho}(x-y)  
- I_{\nu \si}(x-y) I_{\mu \rho}(x-y) ) \,  
\end{gather}
the inversion on $V_F$, the space of antisymmetric
tensor fields, is manifestly conformally covariant as expected since  $F_{\mu \nu}$
is quasi-primary. Under a general gauge transformation we expect
\begin{gather}
\l a_\mu (x) a_\si (y) \r \to \l a_\mu (x) a_\si (y) \r + \pr^x{}_{\!\mu} 
\Lambda_\si(x,y) + {\bar \Lambda}_\mu (x,y) {\overleftarrow \pr}{}^{\!\! y}{}_{\! \si}
\, .
\end{gather}
If we choose in this
\begin{align}
\Lambda_\si(x,y) &= -\frac{1}{4\pi^2} \Big ( \ln(x-y)^2 - \half \ln (x-z)^2
\Big )
\frac{(z-y)_\si}{(z-y)^2 } \, , \nonumber \\
{\bar \Lambda}_\mu (x,y) &= -\frac{1}{4\pi^2}\frac{(z-x)_\mu}{(z-x)^2 } 
\Big ( \ln(x-y)^2 - \half \ln (y-z)^2\Big ) \nonumber \, ,
\end{align}
we get, with $X,Y$ as in (\ref{X}),
\begin{gather}
\l a_\mu (x) a_\si (y) \r_2 = \frac{1}{8\pi^2} \bigg ( (1+\alpha) 
\frac{I_{\mu\si}(x-y)}{(x-y)^2} +4  X_\mu Y_\si \bigg ) \, ,
\label{aacon}
\end{gather}
where the point $z$ here plays the role of a gauge parameter. This reduces
to (\ref{aa}) for $z=\infty$. Since
$ I_{\mu\si}(x-y)/(x-y)^2 = - \half \pr^x{}_{\! \mu} \ln (x-y)^2 
{\overleftarrow \pr}{}^{\!\! y}{}_{\! \si}$ the first term, depending on
$\alpha$, is a pure gauge. The expression in (\ref{aacon})
is conformally covariant if conformal transformations are extended to the
three points $x,y,z$. Similar non-local gauge transformations and
essentially the same result for the photon propagator were described in
\cite{Johnson}.
Using the result (\ref{aacon}) we then have 
\begin{gather}
\l F_{\mu \nu}(x)  a_\si (y) \r_2 = \frac{1}{\pi^2} \frac{ X_{[\mu} 
I_{\nu]\si} (x-y)}{(x-y)^2 } \, ,
\label{Fa}
\end{gather}
which satisfies 
\begin{gather}
\pr_\mu \l F_{\mu \nu}(x)  a_\si (y) \r_2 = \frac{1}{2\pi^2}\, X^2 \bigg (
\frac{I_{\nu\si}(x-y)}{(x-y)^2} - 2 X_\nu Y_\si \bigg ) 
= \frac{1}{2\pi^2} ( X^2 X_\nu) {\overleftarrow \pr}{}^{\!\! y} {}_{\! \si} \, .
\label{Fac}
\end{gather}

With these results, in any gauge, we have
\begin{gather}
\l T_{\mu \nu} (x) T_{\si \rho} (y) F_{\be \ga}F^*{}_{\! \be \ga} (z) \r = 0 \, .
\end{gather}
As a consequence 
$\pr^z{}_{\! \omega}\l T_{\mu \nu} (x) T_{\si \rho} (y) K_\om (z)\r =0$
for $z\ne x,y$. If we
use the covariant  form (\ref{aa}) to compute the three point function
of the current $K_\omega$ with two energy momentum tensors the  photon triangle
diagrams give
\begin{gather}
\l T_{\mu \nu} (x) T_{\si \rho} (y) K_\om (z) \r_1  \hspace{11cm}
\nonumber\\ =  
- \frac{16}{\pi^6} \,
\eps_{\omega \al \be \ga} \E^T{}_{\!\!\!\mu \nu, \mu' \nu'}
\E^T{}_{\!\!\! \si \rho, \si'\rho'}\, \frac{\de_{\al [\mu'}\, 
(z-x)_{\lambda]}}{(z-x)^4} \,
\frac{\I^F{}_{\!\! \nu' \lambda, \si' \kappa}
  (x-y)}{(x-y)^4} \,
 \, \frac{ \I^F{}_{\!\! \rho' \kappa, \beta
    \gamma}(y-z)}{(y-z)^4} \nonumber\\  + \, \{ (x,\mu, \nu)
\leftrightarrow (y,\si, \rho) \} \, ,\hspace{5cm}
\label{photons}
\end{gather}
which is not conformally invariant. However by  using the conformal results
(\ref{FF}) and (\ref{Fa}) we obtain the gauge equivalent expression
\begin{align}
\l T_{\mu \nu} (x) T_{\si \rho} (y) K_\om (z) \r_2
&=  \frac{8}{ \pi^6} \, \frac{\I^T{}_{\!\!\mu \nu, \mu' \nu'}(x-z)
\I^T{}_{\!\!\si \rho, \si' \rho'}(y-z)}{(x-z)^8 (y-z)^8} \, 
\eps_{ \mu' \si' \lambda \omega} \frac{Z_{\lambda}}{Z^6} \left( \delta_{\nu' \rho'}
- 2 \frac{Z_{\nu'} Z_{\rho'}}{Z^2} \right) \, ,
\label{tri2}
\end{align}
which is of the form (\ref{tta}) required for conformal invariance. However
neither (\ref{photons}) nor (\ref{tri2}) satisfy the requirements arising from
$\pr^x {}_{\! \mu}  T_{\mu \nu} (x) =0 ,\ \pr^y {}_{\! \si } T_{\si \rho} (y) =0$,
as shown by the disagreement of (\ref{tri2}) with (\ref{ttta}) and (\ref{ba}). 
This is a consequence of the non-zero result in (\ref{Fac}).\footnote{In the covariant
gauge, with the photon two point function given by (\ref{aa}), the equation of motion 
becomes $\pr_\mu F_{\mu\nu} + \pr_\nu b = 0$ where $b= \pr{\cdot a}/\alpha $. The
associated conserved energy momentum tensor is then given by (\ref{TX})
$X_{\mu\nu} = -2 \pr_{(\mu} {\bar c} \, a_{\nu)} + \de_{\mu\nu}( \pr {\bar c}
{\cdot a} + \frac{1}{2} \alpha \, {\bar c} b ) $, which is not traceless. The
BRS action in this abelian theory is $s a_\mu = \pr_\mu c , \, s c =0 , \,
s  {\bar c} = b , \, s b=0$.  The additional gauge dependent terms in the
energy momentum tensor contribute in the three point function involving the gauge
dependent current $K_\omega$.}
Nevertheless an expression in which energy-momentum is conserved
may be obtained by performing a further
transformation  of the current $K_\omega$ of the form shown in (\ref{Kg}),
so that
\begin{gather}
\l T_{\mu \nu} (x) T_{\si \rho} (y) K_\om (z) \r_3 =
\l T_{\mu \nu} (x) T_{\si \rho} (y) K_\om (z) \r_2 
+ \pr^z{}_{\! \al} W_{\mu\nu, \si \rho, \omega \al }(x,y,z)\, .
\label{TTK}
\end{gather}
Assuming the conformally covariant form 
\begin{gather}
 W_{\mu\nu, \si \rho, \omega \al }(x,y,z)
=   \label{ttw}
\frac{\I^T{}_{\!\!\mu \nu, \mu' \nu'} (x-z) \, \I^T{}_{\!\!\si \rho, \si' \rho'}
(y-z) }{(x-z)^8 (y-z)^8} \, t^{TTW}{}_{\!\!\!\mu'\nu'\si'\rho'\omega\alpha }(Z) \, ,
\end{gather}
with
\begin{gather}
t^{TTW}{}_{\!\!\!\mu\nu\si\rho\omega\alpha }(Z) = {\cal C}\,
\frac{1}{Z^{10}} \E^T{}_{\!\! \mu \nu, \mu' \nu'}
\E^T{}_{\!\! \si \rho, \si' \rho'} \left(\eps_{\omega \al \be \mu'} Z_\be Z_{\nu'}
Z_{\si'} Z_{\rho'} + \eps_{\omega \al \be \si'} Z_\be Z_{\rho'}
Z_{\mu'} Z_{\nu'} \right) \, ,
\end{gather}
which is the unique form satisfying the necessary symmetry condition
\begin{gather}
t^{TTW}{}_{\!\!\!\mu\nu\si\rho\omega\alpha }(Z) =
t^{TTW}{}_{\!\!\!\si\rho\mu\nu\omega\alpha }(-Z) \, ,
\end{gather}
gives
\begin{gather}
\pr^z{}_{\!\al} W_{\mu\nu, \si \rho, \omega \al }(x,y,z) =
4 \, {\cal C} \, \frac{\I^T{}_{\!\!\mu \nu, \mu' \nu'}(x-z) \,
\I^T{}_{\!\!\si \rho, \si' \rho'}(y-z)}{(x-z)^8 (y-z)^8} \, 
\eps_{ \mu' \si' \omega \lambda} \frac{Z_{\lambda} {Z_{\nu'}
    Z_{\rho'}}}{Z^8} \, . \hspace{2.2cm}
\label{g2}
\end{gather}
Hence choosing ${\cal C} = - 8/\pi^6$ leads in (\ref{TTK}) to
\begin{gather} \label{ttaend}
\l T_{\mu \nu} (x) T_{\si \rho} (y) K_\om (z) \r_3
=  \frac{8}{ \pi^6} \, \frac{\I^T{}_{\!\!\mu \nu, \mu' \nu'}(x-z) \,
\I^T{}_{\!\!\si \rho, \si' \rho'}(y-z)}{(x-z)^8 (y-z)^8} \, 
\eps_{\omega \mu' \si' \lambda} \, \frac{Z_{\lambda}}{Z^6} 
\left( \delta_{\nu' \rho'}- 6 \, \frac{Z_{\nu'} Z_{\rho'}}{Z^2} \right) \, , 
\end{gather}                    
which agrees with (\ref{tta}) and with (\ref{ttta})
and so satisfies the energy-momentum conservation condition
(\ref{ba}). By comparing the coefficients we read off that
\begin{gather}
\A = \frac{8}{\pi^6}
\end{gather}
for photons. Since (\ref{ttaend}) is conserved, the anomaly discussion of section 4 above applies to this example as well, 
such that from (\ref{ca}) the anomaly coefficient in
 (\ref{cano}) is
\begin{gather}
c_A = \frac{1}{24 \pi^2}
\end{gather}
for the photon triangle anomaly. This result agrees with the result
of \cite{Dolgov}, according to which the Chern-Simons current 
(with the scale set by (\ref{Kom}))
has an anomaly
\begin{gather}
\nabla_\om \l K^\om \r = \frac{1}{48 \pi^2} \, R^{\mu\nu\si\rho}
R^*{}_{\! \mu\nu\si\rho} \, 
\end{gather}
on curved space, up to a factor of $2$ for the numerical value of 
the coefficient.\footnote{It is difficult to discern the origin for this numerical
discrepancy, since the
curved space calculation of \cite{Dolgov} is very different from the flat space
triangle diagram calculation performed here.} Thus this second example confirms that
there is a unique conserved 
form for the conformal three point function, which is uniquely
related to the axial anomaly present on curved space. 
It should be noted that since the $\pr^z{}_{\! \omega}$ derivative
of (\ref{g2}) vanishes identically, the transformations
performed do not alter the anomaly.

\section{Conclusion}

\setcounter{equation}{0}

We have constructed the three point function involving the axial vector current
as well as two energy-momentum tensors for four dimensional conformal field
theories and shown that this three point function is determined by conformal
invariance up to an overall factor. By analysing the short-distance behaviour
we have related this factor to the coefficient of the gravitational axial
anomaly. 

Our result is of particular interest for supersymmetric theories, for which
it applies to the three point function 
$\l T_{\mu \nu}(x) T_{\si \rho}(y) R_\omega (z) \r$ involving the
axial $R$ symmetry current.
In this case there is a gravitational axial anomaly whose coefficient satisfies
\begin{gather}
c_A \propto c - a \, ,
\end{gather}
with $c$, $a$ the coefficients of the Weyl and Euler densities in the
curved space trace anomaly respectively. 

A direct consequence of our result (\ref{ca}) is that for 
theories for which the
gravitational axial anomaly is absent, $c_A=0$, the three point function 
$\l T_{\mu \nu}(x) T_{\si \rho}(y) A_\omega (z) \r$ vanishes altogether, even 
without taking the divergence with respect to the axial vector leg.           
This applies in particular to the three point function
$\l T_{\mu \nu}(x) T_{\si \rho}(y) R_\omega (z) \r$ for
specific supersymmetric theories for which $c=a$. For such theories a
similar result was
found in \cite{Anselmi} by considering the operator product expansion.
Examples for theories with $c=a$ have been considered recently
in the context of the AdS/CFT correspondence, for instance in the
references listed in  \cite{AdS}.

\vspace{1cm}

{\bf Acknowledgements} \hspace{1em} I am very grateful to Prof.~D.~Z.~Freedman
for pointing out the photon triangle anomaly, and for helpful
advice on this issue.
Furthermore I would like to thank Hugh Osborn for many
useful discussions.

\end{document}